%% file: paper.tex
\documentclass[twocolumn]{aastex63}
\usepackage{verbatim}
\usepackage{comment} 
\usepackage{amsmath}

\usepackage{calrsfs}
\DeclareMathAlphabet{\pazocal}{OMS}{zplm}{m}{n}
\newcommand{\DTG}{\mathcal{D}}

\makeatletter
\def\@hex@@Hex#1%
 {\if a#1A\else \if b#1B\else \if c#1C\else \if d#1D\else
  \if e#1E\else \if f#1F\else #1\fi\fi\fi\fi\fi\fi \@hex@Hex}
\makeatother
\definecolor{afcolor}{HTML}{b3443c}

\revised{\today}

\shorttitle{Dust in Blue Monsters}
\shortauthors{A. Ferrara et al.}
\graphicspath{{./}{figures/}}

\begin{document}
\include{definitions}

\title{Blue Monsters at $z>10$. Where has all their dust gone?}
\correspondingauthor{Andrea Ferrara}
\email{andrea.ferrara@sns.it}
\author[0000-0002-9400-7312]{Andrea Ferrara}
\author[0000-0002-7129-5761]{Andrea Pallottini}
\affil{Scuola Normale Superiore, Piazza dei Cavalieri 7, 50126 Pisa, Italy}
\author[0000-0002-2906-2200]{Laura Sommovigo}
\affil{Center for Computational Astrophysics, Flatiron Institute, 162 5th Avenue, New York, NY 10010, USA}
\
    
\begin{abstract}
The properties of luminous, blue (a.k.a. Blue Monsters), super-early galaxies at redshift $z>10$ have been successfully explained by the attenuation-free model (AFM) in which dust is pushed to kpc-scales by radiation-driven outflows.
As an alternative to AFM, here we assess whether {attenuation-free} conditions can be replaced by a \textit{dust-free} scenario in which dust is produced in very limited amounts and/or later destroyed in the interstellar medium.
To this aim we compare the predicted values of the dust-to-stellar mass ratio, $\xi_d$, with those measured in 15 galaxies at $z>10$ from JWST spectra, when outflows are not included. Our model constrains $\xi_d$ as a function of several parameters by allowing wide variations in the IMF, dust/metal production, and dust destruction for a set of SN progenitor models and explosion energies. We find $\log \xi_d \approx -2.2$ for all systems, indicative of the dominant role of SN dust production over destruction in these early galaxies.
Such value is strikingly different from the data, which instead indicate $\log \xi_d \simlt -4$. 
We conclude that dust destruction alone can hardly explain the transparency of Blue Monsters. Other mechanisms, such as outflows, might be required.
\end{abstract}

\section{Introduction} \label{sec:Intro}

In just two years of operations, the \textit{James Webb Space Telescope} (JWST) has dramatically boosted the exploration of the early universe. One of the major discoveries is a large number of super-early ($z>10$) galaxies 
\citep{Naidu22, Arrabal23, Hsiao23, Wang23, Fujimoto23b, Atek22, Curtis23, Robertson23, Bunker23, Tacchella23, Arrabal23, Finkelstein23, Castellano24, Zavala24, Helton24, Carniani24a, Robertson24} that challenge expectations based on pre-JWST data, theoretical predictions, and perhaps even $\Lambda$CDM-based\footnote{Throughout the paper, we assume a flat Universe with the following cosmological parameters: $\Omega_{\rm M} = 0.3075$, $\Omega_{\Lambda} = 1- \Omega_{\rm M}$, and $\Omega_{\rm b} = 0.0486$,  $h=0.6774$, $\sigma_8=0.826$, where $\Omega_{M}$, $\Omega_{\Lambda}$, and $\Omega_{b}$ are the total matter, vacuum, and baryon densities, in units of the critical density; $h$ is the Hubble constant in units of $100\,\kms$, and $\sigma_8$ is the late-time fluctuation amplitude parameter \citep{planck:2015}.} galaxy formation scenarios. 

Super-early galaxies are characterised by bright UV luminosities ($M_{\rm UV} \simlt -20$), steep UV spectral slopes ($\beta \simlt -2.2$, \citealt{Topping_2022,Cullen_2024,Morales_2024}), compact sizes (effective radius $r_e \approx 200$ pc, \citealt{Baggen_2023,Morishita_2024,Carniani24a}), large (for their epoch) stellar masses ($M_\star \approx 10^9 M_\odot$). 
These properties have earned them the nickname of Blue Monsters \citep{Ziparo23}. Their large comoving number density ($n \approx 10^{-5}-10^{-6}\ \rm cMpc^{-3}$, see e.g. \citealt{Casey_2024,Harvey_2024}) is very hard to explain in the framework of standard, pre-JWST models.

These additions include (a) star formation variability \citep{Furlanetto_22,Mason23,Mirocha23,Sun_2023,Pallottini23}, (b) reduced feedback resulting in a higher star formation efficiency \citep{Dekel23,Li23}, (c) a top-heavy IMF \citep{Inayoshi22,Wang_2023,Trinca24,Hutter_2024}, although see \citet{Cueto23}, and more exotic solutions, as (d) the modification of $\Lambda$CDM via the presence of primordial black holes \citep{Liu23}. 

As a final possibility, we list the so-called \quotes{Attenuation-Free Model} \citep[AFM,][]{Ferrara23a, Ziparo23, Fiore23, Ferrara24a, Ferrara24b} for which bright luminosities and blue colors result from extremely low dust attenuation conditions. 
%
According to AFM, dust is produced by stars in \quotes{standard} net\footnote{That is, after reprocessing by the reverse shock thermalizing the ejecta} amounts (i.e. dust-to-stellar ratios $\xi_d \sim 0.001-0.01$, see e.g. \citealt{Inami22, Sommovigo22, Dayal22, Witstok23,Algera24} for constraints up to $z\sim 7$) and later pushed by outflows on kpc-scales. In this scenario, dust is not destroyed, but simply moved to distances such that the attenuation is decreased to the observed value (for a fixed dust mass, $A_V \propto 1/r^2$, where $r$ is the galactocentric radius).

The solution proposed by AFM is based on the notion that, due to their large specific star formation rates, early galaxies can go through super-Eddington phases in which they launch powerful outflows driven by radiation-pressure onto dust \citep{Ziparo23, Fiore23, Ferrara24a}.
AFM consistently explains the shape of the UV luminosity functions \citep{Ferrara23a}, the evolution of the cosmic SFR density \citep{Ferrara24a}, the star formation history and properties of individual galaxies \citep{Ferrara24b}, induces a star formation mini-quenching \citep{Gelli23}, and makes specific predictions for ALMA targeted observations \citep{Ferrara24c}. 

In spite of this success, it is important to consider alternatives to the AFM. Here instead we postulate that the surviving dust amount is low enough to satisfy the experimental $A_V$ bound. Instead of invoking outflows, this can be achieved by (a) decreasing the net SN yield, or (b) by increasing the destruction rate by interstellar shocks once grains are injected in the interstellar medium (ISM). The physics governing these processes is understood, but it depends on parameters that are only partially constrained \citep{Dwek_98, Bianchi07,Martínez-González_2018,Sarangi_18, Slavin_20, Kirchschlager_24}.

In the present study, our aim is to assess whether \textit{attenuation-free} conditions (AFM) can be soundly replaced by \textit{dust-free} conditions in which dust has been either produced in very limited amounts or destroyed thereafter. To answer this question it is necessary to perform a thorough evaluation of the production/destruction processes during the evolution of the galaxy including an extensive coverage of the free-parameter space. 

We anticipate that our study concludes that dust-free conditions are hardly established without running into severe conflicts with observed physical conditions and observational bounds existing across comic epochs. This makes it difficult to consider (almost) dust-free conditions as a valid alternative to AFM in order to explain the properties of Blue Monsters.

%
%
\begin{table*}
\begin{minipage}{160mm}
\caption{Relevant properties of spectroscopically confirmed super-early galaxies at $z>10$}
\begin{tabular}{cccccccc}
\hline\hline
ID             &redshift&      $A_V$ [mag]        &     $\log(Z/\zsun)$    &      $r_e$ [pc]      &       $\log(M_d/\msun)$         &  $\log(M_\star/\msun)$  &     $\log\xi_d$         \\
(1)            &   (2)  &        (3)              &         (4)            &         (5)          &              (6)                &           (7)           &          (8)            \\
\hline 
CEERS2-7929$^a$& 10.10  & $0.14^{+0.29}_{-0.14}$  & \textemdash            & $520^{+127}_{-127}$  & $5.2^{+10.9}_{-5.2}\times 10^4$ &  $8.50^{+0.30}_{-0.40}$ & $-3.78^{+0.99}_{-0.99}$\\ 
MACS0647-JD$^b$& 10.17  & $< 0.01$                & $-0.90^{+0.09}_{-0.09}$& $70^{+24}_{-24}$     & $<6.7\times 10^1$               &  $7.50^{+0.10}_{-0.10}$ & $<-5.67$               \\ 
UNCOVER-37126$^c$& 10.25& $^*0.18^{+0.14}_{-0.14}$& \textemdash            & $426^{+40}_{-42}$    & $4.5^{+3.5}_{-3.5}\times 10^4$  &  $8.16^{+0.08}_{-0.07}$ & $-3.51^{+0.35}_{-0.35}$\\ 
GS-z10-0$^d$   & 10.38  & $0.05^{+0.03}_{-0.02}$  & $-1.91^{+0.25}_{-0.20}$& $< 62$               & $<2.6\times 10^2$               &  $7.58^{+0.19}_{-0.20}$ & $<-5.16$               \\
GN-z11$^e$     & 10.60  & $0.17^{+0.03}_{-0.03}$  & $-0.92^{+0.06}_{-0.05}$& $64^{+20}_{-20}$     & $9.6^{+3.4}_{-3.4}\times 10^3$  &  $8.73^{+0.06}_{-0.06}$ & $-5.75^{+0.17}_{-0.17}$\\
CEERS2-588$^a$ & 11.04  & $0.10^{+0.11}_{-0.07}$  & $-0.84^{+0.16}_{-0.12}$& $< 477$              & $<3.1\times 10^4$               &  $8.99^{+0.54}_{-0.22}$ & $<4.49$                \\ 
Maisie$^f$     & 11.44  & $0.07^{+0.09}_{-0.05}$  & \textemdash            & $340^{+14}_{-14}$    & $1.1^{+1.4}_{-1.1}\times 10^4$  &  $8.40^{+0.30}_{-0.40}$ & $-4.35^{+0.69}_{-0.69}$\\
GS-z11-0$^d$   & 11.58  & $0.18^{+0.06}_{-0.06}$  & $-1.87^{+0.28}_{-0.18}$& $^\dag77^{+8}_{-8}$  & $1.5^{+0.5}_{-0.5}\times 10^3$  &  $8.67^{+0.08}_{-0.13}$ & $-5.50^{+0.20}_{-0.20}$\\
GHZ2$^g$       & 12.34  & $0.04^{+0.07}_{-0.03}$  & $-1.40^{+0.27}_{-0.24}$& $105^{+9}_{-9}$      & $6.1^{+10.6}_{-6.1}\times 10^2$ &  $9.05^{+0.10}_{-0.25}$ & $-6.27^{+0.80}_{-0.80}$\\
UNCOVER-z12$^c$& 12.39  & $0.19^{+0.17}_{-0.10}$  & $-1.34^{+0.60}_{-0.42}$& $426^{+40}_{-42}$    & $4.7^{+4.2}_{-4.2}\times 10^4$  &  $8.35^{+0.14}_{-0.18}$ & $-3.67^{+0.43}_{-0.43}$\\ 
GS-z12-0$^d$   & 12.63  & $0.05^{+0.03}_{-0.02}$  & $-1.44^{+0.23}_{-0.22}$&$^\dag144^{+15}_{-15}$& $1.4^{+0.9}_{-0.9}\times 10^3$  &  $7.64^{+0.66}_{-0.39}$ & $-4.49^{+0.71}_{-0.71}$\\
UNCOVER-z13$^c$& 13.08  & $0.04^{+0.08}_{-0.03}$  & $-1.57^{+0.35}_{-0.28}$& $309^{+110}_{-74}$   & $5.2^{+10.7}_{-5.2}\times 10^3$ &  $8.13^{+0.11}_{-0.15}$ & $-4.41^{+0.89}_{-0.89}$\\ 
GS-z13-0$^d$   & 13.20  & $0.05^{+0.03}_{-0.02}$  & $-1.69^{+0.28}_{-0.31}$& $< 52$               & $<1.9\times 10^2$               &  $7.95^{+0.19}_{-0.29}$ & $<-5.68$               \\
GS-z14-1$^h$   & 13.90  & $0.20^{+0.11}_{-0.07}$  & $-1.10^{+0.60}_{-0.50}$  & $< 160$              & $<7.0\times 10^3$               &  $8.00^{+0.40}_{-0.30}$ & $<-4.15$               \\
GS-z14-0$^h$   & 14.18  & $0.13^{+0.03}_{-0.03}$  & $-0.75^{+0.03}_{-0.03}$  & $260^{+2}_{-2}$      & $1.2^{+0.3}_{-0.3}\times 10^4$  &  $8.84^{+0.09}_{-0.10}$ & $-4.76^{+0.14}_{-0.14}$ \\
\hline
\label{tab:properties}
\end{tabular}
\end{minipage}
\tablecomments{The measured values are taken from the following works: $^a$\citet{Harikane23, Arrabal23}, Castellano (priv. comm.), $^b$\citet{Hsiao23}, $^c$\citet{Wang23, Fujimoto23b, Atek22}, $^d$\citet{Curtis23, Robertson23}, $^e$\citet{Bunker23, Tacchella23}, $^f$\citet{Arrabal23, Finkelstein23}, $^g$\cite{Castellano24}, $^h$\citet{Carniani24a, Carniani24b, Schouws24, Robertson24}. $^*$The attenuation has been derived from the UV spectral slope, for a Meurer extinction law, i.e. $A_V = 1.99(\beta_{\rm UV}-\beta_{\rm int})$, where $\beta_{\rm int}=-2.3$ \citep{McLure18}. $^\dag$Error not provided. We assume a $\approx 10$\% error.} 
\end{table*}

\section{Data}\label{sec:Data}

Super-early galaxies are characterized by very low dust-to-stellar mass ratios. To support this statement we have collected a sample of currently available spectroscopic data, obtained by \textit{JWST}/NIRspec observations, for 15 galaxies at $z>10$ listed in Tab. \ref{tab:properties}. For each of them, the stellar mass, $M_\star$, has been derived by different authors by fitting the Spectral Energy Distribution. The results are reported in column 8 of the Table; masses are in the range $10^{7.5-9.05}\msun$. The same fitting procedure also yields the V-band dust attenuation, $A_V$ (column 3). The effective stellar radius, $r_e$ (column 5) is instead estimated from JWST/NIRcam photometric data. 

To obtain the dust mass we use the standard formula (see e.g. \citealt{Ferrara24a}) 
\begin{equation}
    M_d = \frac{2\pi q}{\kappa_{\rm UV}} r_e^2 \tau_{\rm UV} = 1.38\times 10^5 \frac{q}{2}\left(\frac{r_e}{100 \rm pc}\right)^2 A_V\ \msun
\end{equation}
where $q=(1,2)$ for (disk, spherical) geometry, $\kappa_{\rm UV}=1.26\times 10^5\ \rm cm^2 g^{-1}$ and $\tau_{\rm UV} = 2.655\times(1.086 A_V)$ are the dust mass absorption coefficient for a Milky Way extinction curve \citep{Weingartner01}, and the optical depth at 1500\AA, respectively; the pre-factor 2.655 for $\tau_{\rm UV}$ accounts for the differential attenuation between 1500 \AA\ and the V-band for the adopted extinction curve.
We conservatively assume spherical geometry ($q=2$) as in this case $M_d$ is maximized for a given measured $A_V$. 

The data implies that $M_d$ for all galaxies is generally very low (see Tab. \ref{tab:properties}), with the dustiest galaxy, CEERS2-7929 showing a scant dust mass of $5.2\times 10^4\ \msun$; other galaxies are virtually dust-free, featuring upper limits as low as $67\ \msun$ (MACS0647-JD). 
Such dust scarcity determines uncommonly low values of the dust-to-stellar ratios, $\xi_d = M_d/M_\star$, shown as blue points in Fig. \ref{fig:Fig02}. The data indicates that $\log \xi_d$ is in the range from $-6.27$ to $-3.51$. For comparison, in Fig. \ref{fig:Fig02} we also show the Milky Way value, $\log \xi_d = -3.2$, and the $\xi_d$ interval measured for 14 REBELS galaxies at $z\approx 7$. Interestingly, galaxies in the REBELS sample \citep{Bouwens22a} have a higher dust-to-stellar mass ratios, $\log \xi_d \approx -2$ \citep{Ferrara22a, Dayal22, Sommovigo22}. This conclusion has been recently supported also those few galaxies having multiple-frequency dust continuum observations \citep{Bakx21, Witstok23, Algera24}. 

Thus, super-early galaxies ($z>10$) have a dramatically lower dust content compared to expectations from known EoR ($z\approx 7$) and local galaxies. To gain some insight into such a fundamental issue, we developed the dust evolution model described below.

\section{Model}\label{sec:Model}

To predict the dust mass of super-early galaxies and their dust-to-stellar mass ratio, we follow \citet{Ferrara24b} and write the evolutionary equations governing these two quantities, along with the analogous ones for the metal, $M_Z$, and gas, $M_g$, mass:  
\begin{eqnarray}
&\dot M_d(t)& = [y_d\nu - {\DTG}(t)(1+\beta+\eta) ] \psi(t),  \label{eq:Md}\\
&\dot M_\star(t)& = (1-R) \psi(t),    \label{eq:Ms}\\    
&\dot M_Z(t)& = \{[y_Z\nu-Z(t)](1-R)-\eta\} \psi(t),  \label{eq:MZ}\\
&\dot M_g(t)& = f_b \dot M_a(t) -[(1-R)+\eta] \psi(t);  \label{eq:Mg}
\end{eqnarray}
The function  $\psi = \epsilon_\star {M_g(t)}{t_{\rm ff}^{-1}}$ denotes the star formation rate (SFR). Such process occurs with an \textit{instantaneous} efficiency, $\epsilon_\star$, per free-fall time, $t_{\rm ff} = \zeta H(z)^{-1}$, where $\zeta = 0.06$ and $H(z)^{-1}$ is the Hubble time \citep[][]{Ferrara23a}. Such efficiency implicitly encapsulates the effects of SN feedback in regulating/suppressing star formation.

In eq. \ref{eq:Md} $y_d$ is the \textit{net}, i.e. including the reverse shock destruction of the newly formed grains, dust yield per supernova (SN) formed per unit stellar mass, $\nu$. Both $y_d$ and $\nu$ depend on IMF and other parameters; we defer their description to Sec. \ref{sec:Parameters}.  

Dust sinks (second term in eq. \ref{eq:Md}, r.h.s.) depend on the dust-to-gas ratio, $\DTG = M_d/M_g$. Dust can be removed from the gas by astration, 
destroyed by SN interstellar shocks, and ejected by outflows. These terms are all proportional to $\psi$, with coefficients $(1, \beta, \eta)$, respectively. The destruction coefficient, $\beta$, is rather uncertain, and its treatment is discussed in Sec. \ref{sec:Parameters}.
As the aim of this work is to test explanations alternative to (radiation-driven) outflows for the observed UV transparency of Blue Monsters, we set the outflow loading factor $\eta=0$ in the following (see \citealt{Pallottini24} for the effect of outflows on the mass-metallicity relation).

Eq. \ref{eq:Md} implicitly assumes that: (a) SNe are the only sources of dust at high-$z$ \citep{Todini01, Lesniewska19, Ferrara22a}; (b) grain growth in the interstellar medium can be neglected as its timescale is comparable to the Hubble time \citep{Ferrara16, Dayal22}. Including grain growth would increase $M_d$, thus exacerbating the discrepancy with observations (see below); hence, our choice is conservative.

Eq. \ref{eq:Ms} expresses the fact that stellar mass increases with the SFR with a proportionality constant $(1-R)$ dependent on the gas return fraction\footnote{We implicitly adopt the Instantaneous Recycling Approximation.} of stars, $R$. In turn, $R$ depends on IMF as we will detail in Sec. \ref{sec:Analytical}.   

The last two equations (eqs. \ref{eq:MZ}$-$\ref{eq:Mg}) describe the evolution of the metal and gas mass, respectively. The quantity $y_Z$ is the metal yield per SN; similarly to the dust yield, it depends on the IMF (Sec.  \ref{sec:Parameters}). Depending on metallicity, $Z=M_Z/M_g$, a fraction of the heavy elements is swallowed by star formation. The gas mass increases at a rate $f_b M_a$, where $f_b=\Omega_b/\Omega_m$ is the cosmological baryon fraction, and the dark matter halo growth is fed by cosmological accretion, whose rate (in $\msunyr$) is obtained from numerical simulations, e.g. \citet{Correa15}:
\begin{equation}
\dot M_a(M,z) = 102.2\, h [-a_0 -a_1 (1+z)] E(z) M_{12}; \label{eq:Macc}  
\end{equation}
the parameters $(a_0 , a_1 )$ depend on the halo mass $M(z)$, cosmology and the linear matter power spectrum, and provided\footnote{For reference, when averaged over $8 < \log (M_0/\msun) < 14$, $(a_0 , a_1 )=(0.25, -0.75)$.} in Appendix C of \citet{Correa15}; $E(z) = [\Omega_m(1+z)^3+\Omega_\Lambda)]^{1/2}$, and $M_{12}= M/10^{12} \msun$.
Finally, gas is consumed by star formation and partly returned to the interstellar medium (last term in eq. \ref{eq:Mg}).   

For later use, it is convenient to rewrite the accretion term $f_b \dot M_a$ as $a(z) \psi$, with $a(z)$ to be defined\footnote{We compute $a(z)$ at the observed galaxy redshift as the star formation histories of $z>10$ galaxies are short ($20-30$ Myr). Hence, we neglect the small change of $\dot M_a$ in that time frame.}. To this aim we, use the expression of $\psi$ given in eq. \ref{eq:Mg} to substitute for halo mass in eq. \ref{eq:Macc}, finally obtaining
\begin{equation}
a(z, \epsilon_\star) = 6.23[-a_0 -a_1 (1+z)]\left(\frac{0.01}{\epsilon_\star}\right); \label{eq:a}  
\end{equation}
for reference, $a(z=10)\simeq 50 (0.01/\epsilon_\star)$, i.e. the accretion rate largely exceeds the SFR (unless $\epsilon_\star \simgt 0.5$). In these conditions, the gas fraction $f_g = M_g/(M_g + M_\star) \simeq 1$, and $M_g \approx f_b M$.

\section{Analytical insights}\label{sec:Analytical}

While is it possible to solve the evolutionary eqs. \ref{eq:Md}-\ref{eq:Mg} numerically as done in \citet{Ferrara24b}, to gain a deeper physical insight on dust enrichment we seek here for analytical solutions.
To fix ideas, in this Section for numerical estimates we adopt fiducial values for the five parameters of the model. These are: $\nu^{-1} = 52.89 M_\odot$, $R=0.61$, $y_d=0.1 M_\odot$, $y_Z=2.41 M_\odot$, and $\beta = 4.7$. A detailed discussion of these parameters, and their uncertainties is given in Sec. \ref{sec:Parameters}; the effects on the final results of their combined variation are presented in Sec. \ref{sec:Results}.  

\subsection{Are SNe predominantly dust sources or sinks?}\label{sec:Sinks}

First, we note from eq. \ref{eq:Md} that SNe act both as dust sources and sinks. Dust is produced immediately (few hundreds days) after the explosion in the expanding SN envelope, where it is also processed by the reverse shock thermalising the ejecta. Behind the forward shock propagating in the ISM, pre-existing dust is partially destroyed by sputtering and shattering. 

To assess whether SNe are net dust producers or destroyers, let us have a closer look at eq. \ref{eq:Md}. By comparing the two competing terms, we conclude that SNe are net dust producers as long as    
\begin{equation}
\DTG < \DTG_{\rm crit} \equiv \frac{y_d\nu}{(1+\beta)} \approx 0.05 \DTG_{\rm MW}. \label{eq:Dcrit}  
\end{equation}
For the numerical estimate we have used the fiducial values of the parameters; the adopted value of the Milky Way dust-to-gas ratio is $\DTG_{\rm MW} = 1/162$ \citep{Remy14}. Thus, in the early evolutionary stages, when $\DTG$ is low, SNe are net dust sources. In more evolved galaxies they predominantly destroy the dust created by other sources (AGB and late-type stars) or processes (such as grain growth). 

We pause to emphasize an important fact. A key difference between dust destruction ($\beta$) and outflows ($\eta$) is that while sputtering and shocks destroy dust, outflows simply carry it away from stars. Hence, the value of $\xi_d$ depends on the physical scale on which it is measured. In other words, although the $\beta$ and $\eta$ enter eq. \ref{eq:Md} as an additive term, destruction and outflows do not have the same physical consequences.

We can also translate the above condition into one on metallicity, by adopting a $\DTG - Z$ relation, which we take\footnote{\citet{Remy14} give also an alternative broken power-law fit (see their Table 2) from which we find a similar result, $Z_{\rm crit} = 0.12 Z_\odot$} from \citep[][see also \citealt{Lisenfeld98}]{Remy14}. These authors find $\DTG = (Z/Z_\odot)^{1.62}\ \DTG_{\rm MW}$, where $Z_\odot=0.0142$ \citep{Asplund09}. Hence, $\DTG_{\rm crit} = 0.05 \DTG_{\rm MW}$ would correspond to $Z_{\rm crit} = 0.16 Z_\odot$. From Tab. \ref{tab:properties} we see that none of the super-early galaxy has $Z > Z_{\rm crit}$. We conclude that their dust is consistently produced by SNe, and inefficiently destroyed in the ISM. 

\subsection{A closed-form solution}

To obtain a closed-form solution for $\xi_d$, let us consider eq. \ref{eq:Md} and eq. \ref{eq:Ms}. With simple algebraic manipulations, and using the definition of $\DTG_{\rm crit}$ in eq. \ref{eq:Dcrit}, we find that the dust-to-stellar mass ratio can be written as
\begin{eqnarray}\label{eq:xid}
\xi_d &=& \frac{1}{y} \left[ 1 - \exp{(-\xi_d^0 y})\right],\\
y &=& \frac{1}{\DTG_{\rm crit}} \frac{M_\star}{M_g}. 
\end{eqnarray}
We have introduced the production-only dust-to-stellar mass ratio $\xi_d^0 = y_d \nu/(1-R)$, the value $\xi_d$ would attain if dust would not be post-processed in the ISM by astration or destruction (or growth). This is the maximum value achievable by $\xi_d$. 

We first notice that in the limiting case $M_\star/M_g \ll 1$ (gas-dominated systems), using the fiducial values of the parameters, $\log \xi_d \simeq \log \xi_d^0 = -2.2$, a value $\ge 100$ higher than inferred for super-early galaxies (Fig. \ref{fig:Fig02}). 

A general use of eq. \ref{eq:xid} requires instead an estimate of $M_\star/M_g$. As $M_\star$ is known for these galaxies, to make progress we need to constrain $M_g$. At the moment, detailed information on the gas content of $z>10$ galaxies is lacking. This will have to await for ALMA observations of these systems; in particular, the [CII]158$\mu$m line appears to be the best gas tracer \citep{Zanella18}.  

Fortunately, though, we can use the metallicity data\footnote{For the three galaxies for which $Z$ measurements are not available (see Tab. \ref{tab:properties}) we use the mean value of the sample.} provided by the exquisite {\it JWST} spectra to obtain $M_g$. To this aim, consider the other two equations (eqs. \ref{eq:MZ} and \ref{eq:Mg}) in the system. With simple algebra it is easy to show that the gas metallicity of a purely accreting, star forming galaxy with no outflows ($\eta=0$) is 
\begin{equation}
Z = y_Z\nu  \frac{(1-R)}{a(z,\epsilon_\star)} \left[1-\left(\frac{M_g}{M_0} \right)^{a/q}\right] \simeq  y_Z\nu  \frac{(1-R)}{a(z,\epsilon_\star)}.
\label{eq:Z}  
\end{equation}
Here we have introduced $q = 1-R-a \approx -a$ as $a \gg (1-R)$ (see eq. \ref{eq:a}); hence $a/q \approx -1$. The last equality comes instead from the fact that the gas mass at the final redshift is much larger than that available when star formation begins, $M_0$. Canonically, this epoch coincides with the moment at which the halo growth crosses the value $M \simeq 10^8 M_\odot$, marking the boundary of the atomic cooling regime (virial temperature $T> 10^4$ K). 

By noting\footnote{In general, by time integration of eq. \ref{eq:Ms}, one would obtain 
\begin{equation}
M_\star = (1-R)\int \epsilon_\star M_g t_{\rm ff}^{-1} dt = (1-R)\langle \epsilon_\star \rangle M_g,
\end{equation}
where $\langle \epsilon_\star \rangle$ is the gas conversion efficiency. If star formation lasts for about a free-fall time ($\approx 30$ Myr at $z=10$), than $\langle \epsilon_\star \rangle\approx \epsilon_\star$. Therefore, we neglect the small difference, and simply write $M_\star = (1-R)\epsilon_\star M_g$.} that $M_\star = (1-R)\epsilon_\star M_g$, and using the result obtained in eq. \ref{eq:Z} we can write
that 
\begin{equation}
\frac{M_g}{M_\star} = \frac{1}{(1-R)\epsilon_\star} = \frac{2 y_Z \nu}{Z}, 
\label{eq:MgMd}    
\end{equation}
having also used $a(z=10,\epsilon_\star) \simeq 1/2\epsilon_\star$ (see eq. \ref{eq:a}).

To obtain the small observed values, $\xi_d \simeq 10^{-5}$, the quantity $y = {\DTG_{\rm crit}}^{-1} ({M_\star}/{M_g}) = Z/2 y_Z \nu {\DTG_{\rm crit}}$ in eq. \ref{eq:xid} has to be large. In this case the equation simplifies and becomes (using fiducial values of the parameters)  
\begin{equation}
    \xi_d \simeq 2 \DTG_{\rm crit} \frac{y_Z\nu}{Z} = 6.4 \DTG_{\rm crit} \left(\frac{Z_\odot}{Z}\right).
\end{equation}
Hence, imposing $\xi_d \simeq 10^{-5}$, and given the average metallicity of the sample $Z\simeq 0.05 Z_\odot$, it should be $\DTG_{\rm crit} \simeq 8\times 10^{-8}$. This low $\DTG_{\rm crit}$ value would require an unphysically high value $\beta\simeq 2.3\times 10^4$. Such destruction efficiency is orders of magnitude higher than predicted by all dust destruction models and observations, as discussed in Sec. \ref{sec:Parameters}. 

\subsection{Star formation efficiency}\label{sec:efficiency}

We conclude this Section by emphasizing that high star formation efficiencies necessarily imply large metallicities (see eq. \ref{eq:Z}). For the fiducial value $y_Z= 2.41 M_\odot$, we find that at $z=10$, $Z = 0.02 (\epsilon_\star/0.01) Z_\odot$. Hence, the metallicity of all galaxies in Tab. \ref{tab:properties} can be matched with small variations of the efficiency in the range $0.01-0.05$. 
Larger star formation efficiencies would dramatically overproduce the predicted metallicities with respect to the observed ones. We recall that by choice the scenario investigated here does not include outflows.

We also note that, as $a(z, \epsilon_\star)$ depends linearly on redshift, the metallicity grows with time. For example, the same calculation performed at $z=2$, yields $Z= 0.08 (\epsilon_\star/0.01) Z_\odot$.

\section{Model parameters}\label{sec:Parameters}

Eq. \ref{eq:xid} contains a number of parameters we allow to vary. These are: $\nu,\ R,\ y_d,\ y_Z$ and $\beta$. So far, for our estimates we have used fiducial values for these parameters, but it is important to assess the impact of their combined variation.
As already mentioned, the first four depend on the IMF and on the SN progenitor properties (rotation, explosion energy); the dust yield $y_d$ also depends on the ISM gas density affecting the reverse shock destruction efficiency. Finally, $\beta$ controls dust destruction by shocks in the ISM. We discuss how we model these processes in the following.

\subsection{Initial Mass Function}\label{sec:IMF}
We assume that stars form with masses in the range $(m_l, m_u) = (0.1, 100) M_\odot$ according to a Larson IMF \citep{Larson98} normalized to unity in the above range\footnote{Such a choice neglects the possible presence of pair instability SNe, which occur only in the range $140-260\ \msun$ and possibly only at very low metallicity; therefore they are very rare. Moreover, they would further enhance dust and metals yields compared to core-collapse SNe, worsening the problem at hand.}, which follows a Salpeter-like power law at the upper end but flattens below a characteristic stellar mass:
\begin{equation}
     \phi(m) \propto m^{-(1+\alpha)} e^{-m_c/m},   
\label{eq:IMF}
\end{equation}
where $\alpha=1.35$. We take $m_c$ to be a random variable log-normally distributed with mean and standard deviation $(\mu_1, \sigma_1) = (0, 0.5)$ (an example of a random distribution is shown in Fig. \ref{fig:Fig01}). Assuming that stars with $m > 8 M_\odot$ explode as SNe, we derive the number of SNe per unit stellar mass formed, $\nu$, as a function of $m_c$. 

Imposing the Instantaneous Recycling Approximation \citep[][]{Maeder92}, for which stars with mass $m < m_0$ ($m\geq m_0$) live forever (die immediately) the return fraction, $R$, is 
\begin{equation}
     R = \int_{m_0}^{m_u} (m-w_m) \phi(m) dm.   
\label{eq:R}
\end{equation}
We fix $m_0=2.3 M_\odot$, the mass of a star whose lifetime is equal to the cosmic age (483 Myr) at $z=10$, and assume \citep{Weidemann00} that the remnant mass is (a) $w_m = 0.444 + 0.084 m$ for $m<8 M_\odot$; (b) $w_m = 1.4\ M_\odot$ (neutron star) for $8\ M_\odot < m < m_{\rm BH}=40 \ M_\odot$, (c) $w_m = m$ for $m > m_{\rm BH}$ when they collapse into a black hole.

\subsection{Dust yields}\label{subsec:dust_yield}
To compute the dust, $y_d$ yields from SNe, it is necessary to define the stellar progenitor model. In general, the nucleosynthetic yields depend on the explosion energy and on the possible rotation of the star.
We follow \citet[][see their Tabs. 9-12]{Marassi19} who consider 4 different models. The first two models assume a fixed explosion energy of $10^{51}$ erg, and the star is either non-rotating (FE-NR) or rotating (FE-ROT) at a velocity of $300\, \kms$. Instead, the second set adopts a variable explosion energy calibrated to reproduce the amount of $^{56}$Ni obtained from the best fit to the observations for non-rotating (CE-NR) or rotating (CE-ROT) progenitors\footnote{We note that the black hole mass formation is $m_{\rm BH}=40 \msun$ for the standard FE-NR model, but it is larger for the other three models ($m_{\rm BH}=60 \msun$). This is taken into account in the calculation.}. For all cases, the metallicity is fixed to $0.01 Z_\odot$. \citet{Marassi19} perform detailed dust nucleation calculations and obtain $y_d$ for the four models above for 10 different progenitor masses in the range $8-120\  \msun$. The computed dust yields are in the range $\approx 0.1-1.4\ \msun$, with the CE-ROT predicting the highest values (with the notable exception of the $60 \msun$ star producing up to $7.4 \msun$ of dust  in this case).
In our sampling of the parameter space, the stellar model is set by assuming (FE-NR, FE-ROT, CE-NR, CE-ROT) are distributed as a uniform discrete random variable.

\citet{Marassi19} did not consider the processing of the newly formed grain by the reverse shock thermalising the ejecta. We therefore augment their model with the results by \citet{Nozawa07}, who computed the surviving dust mass fraction after the reverse shock, $f_{\rm rev}$ for a range of masses from 13 to 200 $\msun$ and for a range of environmental gas densities in which the explosion takes place, from $n_g = 0.1$ to 10 $\cc$. We linearly interpolate their results on our progenitor masses and gas densities. We further assume the unmixed ejecta case. We point out \citet{Nozawa07} results are strictly valid for the FE-NR case only; lacking more general computations we adopt them for all the four stellar progenitor models.   

%
%
\begin{figure}
\centering\includegraphics[width = 0.48 \textwidth]{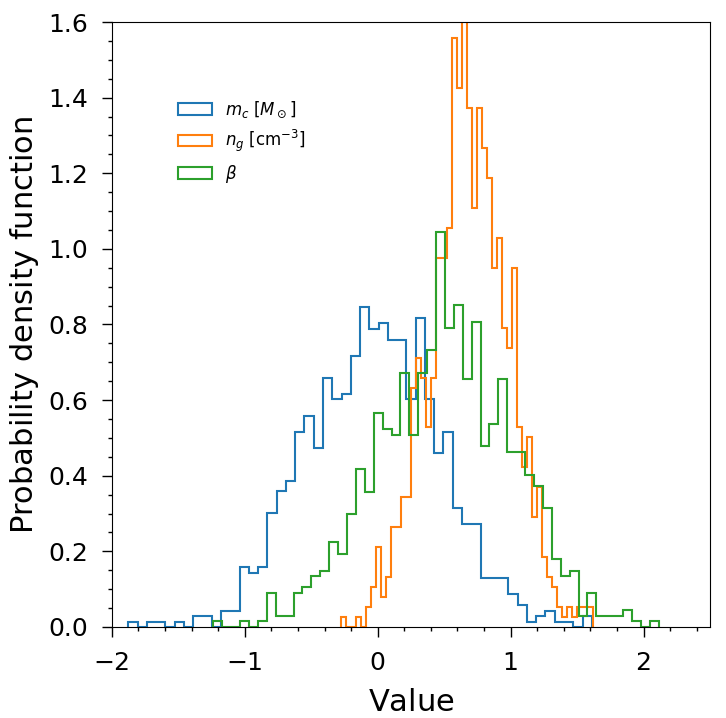}
\caption{Example of a probability density function for three parameters of the problem, the IMF characteristic mass, $m_c$, the ISM gas density, $n_g$, and the destruction coefficient, $\beta$. These are discussed in Sec. \ref{sec:IMF}, Sec. \ref{subsec:dust_yield}, Sec. \ref{sec:destr}, respectively.} 
\label{fig:Fig01}
\end{figure}

We take $n_g$ to be a random variable log-normally distributed with mean and standard deviation $(\mu_3, \sigma_3) = (0.5, 0.5)$ (see Fig. \ref{fig:Fig01}). It is important to note that the mean gas density of early galaxies is thought to be larger than locally \citep{Isobe23}, reaching values of $\approx 10^3 \cc$ at $z\simgt 10$. Nevertheless, the explosion is always going to occur in a relatively low ($1-10\, \cc$) density medium.

To see this, assume that at the star birth the gas density is $n_g^0 = 10^3\ \cc$. Due to the strong ionising power of the massive ($m > 8\ \msun$) SN progenitor, an \HII region will form. The Str{\"o}mgren radius will be filled in a recombination time, or $(n_g^0 \alpha_B)^{-1} \simeq 120$ yr, where $\alpha_B$ is the case-B recombination coefficient of H-atoms. From that moment, the overpressurized \HII region (temperature $T_i\approx 2\times 10^4$ K) will start to expand to restore pressure equilibrium with the surrounding gas.
The external gas temperature in an almost metal-free gas is set by molecular hydrogen to $T_0 \simlt 300$ K; if metallicity is higher, the gas could be even colder. Then the expansion will decrease the density by a factor $T_i/T_0$, bringing it down to $n_g \simlt 15\ \rm cm^{-3}$. 

This simple estimate is in line with radiative transfer simulations (e.g. \citealt{Whalen04}) of HII regions around massive stars. These authors find that even starting from initial gas densities well in excess of $10^3\ \cc$, the HII regions produce an almost uniform distribution of the gas around the stars. They find (see their Fig. 3) that the density in the HII region is $\approx 0.1\ \cc$ after $2.2$ Myr, i.e. approximately the time when the most massive stars explode as SNe. This justifies our choice for the mean value, $\mu_3$ of the gas density, which we allow nevertheless to vary considerably to catch the uncertainties on the gas and stellar properties. 

As a reference, according to \citet{Nozawa07} results, the surviving dust mass fraction for $n_g \approx 1\ (10)\ \cc$ varies from 20 to 47 (4 to 10) percent in the mass range $8\ M_\odot < m < 40 \ M_\odot$. We perform a 2D linear interpolation in the stellar mass -- gas density plane to derive $f_{\rm rev}$ when sampling our model. 

%
%
\begin{figure*}
\centering\includegraphics[width = 1.0 \textwidth]{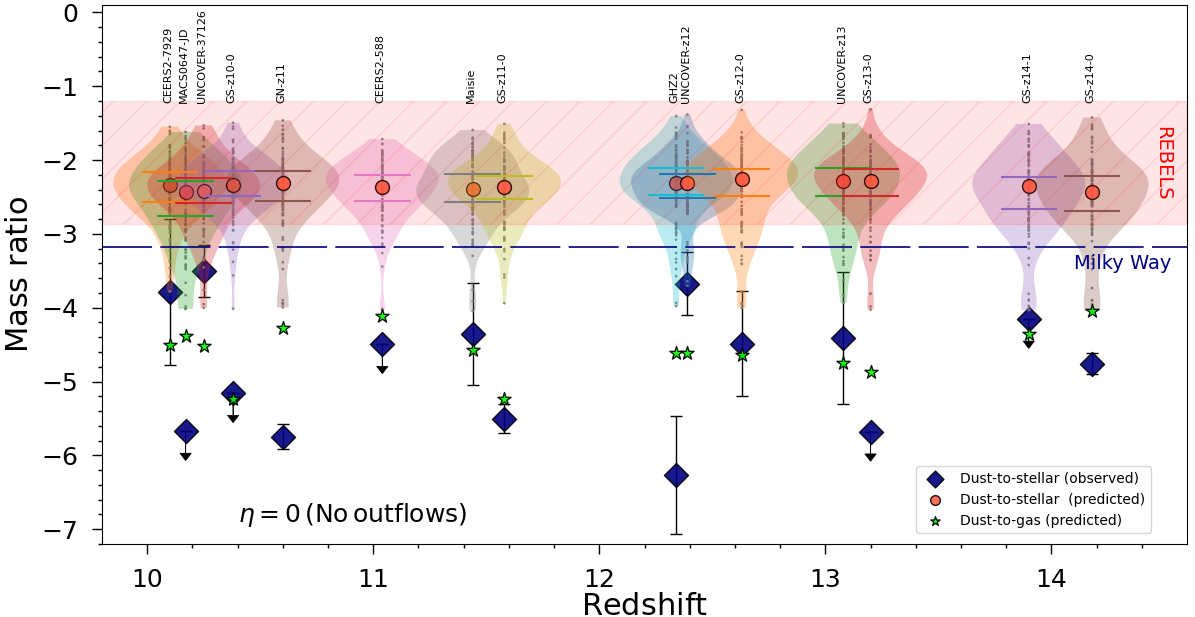}
\caption{Violin plots showing the posterior distribution of $\xi_d$ resulting from the random sampling of the parameter space (see details in Sec. \ref{sec:Parameters}) for 15 super-early galaxies whose redshift is indicated on the horizontal axis and name is shown at the top.
For each galaxy, the red point represents the distribution median value; the horizontal bars inside the violin indicate the 32-nd and 68-th percentiles. The predicted $\xi_d$ values are for $\eta=0$, i.e. when no galactic outflows are allowed. The blue diamonds are the observed dust-to-stellar mass ratios; the green stars are predicted dust-to-gas ratio, $\DTG$. The horizontal hatched red band shows the measured $\xi_d$ interval for 14 REBELS galaxies at $z\approx 7$ (Tab. 1 in \citealt{Ferrara22a}; see also \citealt{Dayal22, Sommovigo22, Inami22}).} 
\label{fig:Fig02}
\end{figure*}

\subsection{Metal yields}
To compute the metal yields, we follow \citet{Kim14} who use the results by \cite{Woosley07}. In practice, the oxygen and iron mass can be converted into a total metal mass using the following formula: 
\begin{equation}
M_Z = 2.09 M_{\rm O} + 1.06 M_{\rm Fe},
\label{eq:Zyield}
\end{equation}
where
\begin{eqnarray}
M_{\rm O} = 0.375 \exp (-17.94/m)\ \msun, \\
M_{\rm Fe} = 27.66 \exp(-51.81/m)\ \msun.
\end{eqnarray}
Eq. \ref{eq:Zyield} is then integrated over the IMF to obtain the metal yield $y_Z$, the average metal mass produced per SN. In the fiducial case, $y_Z=2.41\ M_\odot$; this result is very weakly dependent on metallicity of the stars. We warn, however, that \citet{Woosley07} results have been computed for the standard case FE-NR, and we are applying them also to the other three progenitor models adopted here.  

\subsection{Dust destruction in interstellar shocks}\label{sec:destr}
In addition to reverse shocks discussed in Sec. \ref{subsec:dust_yield}, grains can be destroyed in interstellar shocks produced by SNe (see Sec. \ref{sec:Sinks}) for a discussion of the required conditions). As a result the destruction term is proportional to the star formation rate, $\psi$, via a destruction coefficient $\beta$ (eq. \ref{eq:Md}). The coefficient is itself the product of different terms \citep[e.g.][]{Dayal22}:
\begin{equation}
\beta = \delta_{\rm SN} f_{\rm eff} \nu M_s^{\rm dest}.
\label{eq:beta}
\end{equation}
In the previous eq. $\nu M_s^{\rm dest}$ is the mass of gas in which dust is destroyed by the SN shocks per unit stellar mass formed. A value of $M_s^{\rm dest}= 1603\ \msun$ has been estimated by \citet{Slavin15} by assuming a 50\%+50\% mixture of silicate and carbonaceous grains. \citet{Bocchio14} found instead a $\approx 2.6$ times larger value (for silicates) due to the fact that they assume a different grain destruction efficiency as a function of the shock velocity. 

The other two factors in eq. \ref{eq:beta} represent the fraction of SNe exploding outside the galaxy main body, $\delta_{\rm SN} \simeq 0.36$, and the fraction of warm ($\approx 8000$ K) to hot gas in the galaxy, $f_{\rm eff} \simeq 0.43$ \citep{Slavin15}. These quantities have been calibrated locally \citep{Schneider24} and therefore are somewhat uncertain. At face value they would give a fiducial value $\beta = 4.7$. In view of these uncertainties, we take $\beta$ to be a random variable log-normally distributed with mean and standard deviation $(\mu_4, \sigma_4) = (0.7, 0.3)$ (see Fig. \ref{fig:Fig01}).
    
To summarise, our model contains five parameters that we allow to vary to fully sample the physical uncertainties related to their value in early galaxies. These are $\nu, R, y_d, y_Z, \beta$. For each galaxy we run $N=1000$ models\footnote{We have checked that the posterior distribution of $\xi_d$ does not change significantly if $N$ is increased.}, each for a random combination of the parameters, thus fully sampling the parameter space determining the value of the dust-to-stellar mass ratio obtain in eq. \ref{eq:xid}. As seen from, Fig. \ref{fig:Fig01}, each parameter can take values in a wide range spanning $2-3$ orders magnitude around the mean. We recall that in addition to these parameters, we consider 4 randomly selected different stellar models for the SN progenitor (see Sec. \ref{subsec:dust_yield}).

\section{Results}\label{sec:Results}
The results of our study and condensed in Fig. \ref{fig:Fig02}. There we show the violins plot for the posterior distribution of $\xi_d$ resulting from the random sampling of the parameter space for each of the 15 super-early galaxies. The red point in each distribution represents the median value; also shown with horizontal bars inside the violins are the 32-th and 68-th percentiles.

The median values for all galaxies are rather similar and fall in a narrow range around $\log\xi_d = -2.2$. This fact is indicative that the value of the dust-to-stellar ratio is largely determined by the net (i.e. including the reverse shock destruction) SN dust yield, with destruction in interstellar shocks being negligible.
This result confirms the conclusion drawn in Sec. \ref{sec:Sinks}, where show that for $\DTG < \DTG_{\rm crit} \approx 3\times 10^{-4}$ SNe largely act as dust sources rather than sinks. Indeed, the values we predict for the dust-to-gas ratio, $\DTG$, in $z>10$ systems (green stars in Fig. \ref{fig:Fig02}) are consistently below $10^{-4}$.
The distributions are somewhat skewed towards low $\xi_d$, with some model combinations yielding $\xi_d \approx 10^{-4}$ for most extreme values of the parameters.  

The difference between the predicted $\xi_d$ values and the measured ones (blue diamonds) is striking. Apparently, no combination of parameters, although allowed to vary in a large range around the fiducial values given in the literature, yields $\xi_d < 10^{-4}$, while the majority of that data points are well below that threshold.
Three galaxies (CEERS2-7929, UNCOVER-37126 and UNCOVER-z12) have a measured $\xi_d > 10^{-4}$, but are yet several r.m.s. away from the predicted median. The discrepancy is even more impressive for the most well-studied galaxies as GHZ2, GS-z11-0, and GS-z13-0 for which exquisite-quality spectra are available. These systems show dust-to-stellar ratios as low as $\xi_d \approx 10^{-6}$.

Thus, it appears that the blue colours (steep UV slopes) and transparency (low $A_V$) of super-early galaxies imply extremely low dust-to-stellar mass ratios that cannot be reproduced by the “standard” dust physics (in the absence of outflows) even allowing for a large and simultaneous variation range of all the, admittedly uncertain, dust production and destruction parameters. One could be tempted to dramatically change the dust yields or making the dust destruction extremely efficient. However, going down that path we would be faced with the difficulty of explaining a large body of data for local galaxies, and their intermediate- and high-$z$ counterparts (such as the REBELS galaxy sample at $z\approx 7$), which are well explained by the present model. 

\section{Summary and Discussion}
As a test of the ‘Attenuation-Free Model’ (AFM), we have compared the predicted values of the dust-to-stellar mass ratio, $\xi_d$, measured in fifteen $z>10$ galaxies with those derived from JWST spectra. Differently from AFM, the model considered here does not rely on outflows to remove dust from the galaxy main body. Instead, it constrains the value of $\xi_d$ as a function of various parameters (discussed in detail in Sec. \ref{sec:Parameters}) by allowing variations in the IMF, dust and metal production and dust destruction for a set of SN progenitor stellar models. 

The key results are shown in Fig. \ref{fig:Fig02}, from which we see that the predicted $\xi_d$ range is around $\log\xi_d = -2.2$ for all galaxies, indicative of the dominant role of SN dust production over destruction in these early systems. The predictions are strikingly different from the data, which instead show values $\log \xi_d \simlt -4$. We conclude that negligible dust production or strongly enhanced destruction can hardly reconcile theory and observations without conflicting with well-studied $z\approx 7-8$ galaxies. Other mechanisms, such as outflows, might be required by this evidence. 

In this study we have purposely not allowed for the occurrence of outflows, although cosmological infall has been included in our master eqs. \ref{eq:Md}$-$\ref{eq:Mg}. Our aim has been to challenge the AFM by proposing a radically different solution to the overabundance and colors of bright super-early galaxies.

The AFM scenario \citep{Ferrara24a, Ferrara24b} suggests that if dust could be lifted out of the main galaxy body and into its halo on kpc-scales, the low $A_V$ values can be readily explained by the $1/r^2$ decrease of the dust column density. This hypothesis is supported by the fact that towards high-$z$ outflows become more common as a result of the increased compactness of galaxies and of their on average larger sSFR. Both these conditions favor powerful radiation-driven outflows clearing the dust \citep{Ziparo23, Fiore23, Ferrara24a}. 

In AFM, dust “standard” formation/evolution physics still holds, but dust is dispersed by outflows mimicking a lack of dust, and producing the low observed attenuation (see the case of GS-z14-0 in \citealt{Ferrara24b}). Thus, a no-outflow model as the one presented here, falls far-cry short to reproduce the data, which instead are well reproduced by the AFM. 

Interestingly, the larger and more evolved REBELS galaxies ($z\approx 7$) are showing $\xi_d$ values of the same order ($\approx 1/100$) as predicted here \citep[][]{Ferrara22a, Dayal22, Algera24}. This might indicate that the processes included in this study — in the absence of powerful outflows — continue to regulate the dust abundance for a long cosmic time stretch. Intriguingly, their measured metallicities and gas fractions can be generally explained by a no-outflow model (Algera et al., in prep.), which entails the high $\xi_d$ predicted here. 

It is worth noticing that we have neglected dust growth. Although likely not important at these early redshifts due to the associated long timescales \citep{Dayal22} or prevailing physical conditions \citep{Ferrara16}, dust growth would exacerbate the problem by increasing $\xi_d$ even above the values found here. 

On more general grounds, we note that Blue Monsters are characterized by a sustained SFR around $10-20\ \msunyr$. It is well established \citep[e.g.][]{McKee07, Glover12, Pallottini22} that SF can only occur if the star forming gas is shielded from UV radiation by dust. It is therefore hard to understand how such an active star formation phase could have started and supported in a virtually dust-free environment. 

We conclude with a remark. This work suggests that the dust we observe in Blue Monsters retains the properties of freshly produced one by SN without significant reprocessing in the ISM. If correct, dust formation models \citep{Todini01, Nozawa07, Hirashita15} predict that these readily produced grains should be large, with sizes in the range $0.1-0.5\ \mu$m. In this case we do expect that the extinction curve should become almost frequency-independent (i.e. “gray”). This has been indeed recently found by \citet{Markov24} using a sample of $\approx 100$ JWST galaxy spectra in the redshift range $z=2-11$. Similar conclusions are reached by \citet{Langeroodi24_b}.
%


\acknowledgments
We thank S. Carniani, M. Castellano, Y. Harikane, A. Inoue for useful discussions, data and comments.
This work is supported by the ERC Advanced Grant INTERSTELLAR H2020/740120. 
This research was supported in part by grant NSF PHY-2309135 to the Kavli Institute for Theoretical Physics (KITP). 
Plots in this paper produced with the \textsc{matplotlib} \citep{Hunter07} package for \textsc{PYTHON}.    

\bibliographystyle{aasjournal}
\bibliography{paper}

\end{document}

%% file: definitions.tex


\def\be{\begin{equation}}
\def\ee{\end{equation}}
\newcommand\code[1]{\textsc{\MakeLowercase{#1}}}
\newcommand\quotesingle[1]{`{#1}'}
\newcommand\quotes[1]{``{#1}"}
\def\gsim{\lower.5ex\hbox{\gtsima}} 
\def\lsim{\lower.5ex\hbox{\ltsima}} 
\def\gtsima{$\; \buildrel > \over \sim \;$} 
\def\ltsima{$\; \buildrel < \over \sim \;$} \def\gsim{\lower.5ex\hbox{\gtsima}} 
\def\lsim{\lower.5ex\hbox{\ltsima}} 
\def\simgt{\lower.5ex\hbox{\gtsima}} 
\def\simlt{\lower.5ex\hbox{\ltsima}}

\def\msun{{\rm M}_{\odot}}
\def\lsun{{\rm L}_{\odot}}
\def\dsun{{\cal D}_{\odot}}
\def\fsun{\xi_{\odot}}
\def\zsun{{\rm Z}_{\odot}}
\def\msunyr{\msun {\rm yr}^{-1}}
\def\gdens{\msun\,{\rm kpc}^{-2}}
\def\sfrdens{\msun\,{\rm yr}^{-1}\,{\rm kpc}^{-2}}

\def\mum{\mu {\rm m}}
\newcommand{\angstrom}{\mbox{\normalfont\AA}}
\def\cc{\rm cm^{-3}}
\def\uflux{{\rm erg}\,{\rm s}^{-1} {\rm cm}^{-2} }

\def\fdust{\xi_{d}}
\def\fesc{f_{\rm esc}\,}
\def\td{\tau_{sd}}
\def\Sg{$\Sigma_{g}$}
\def\S*{$\Sigma_{\rm SFR}$}
\def\Ssfr{\Sigma_{\rm SFR}}
\def\Sgas{\Sigma_{\rm g}}
\def\Sstar{\Sigma_{\rm *}}
\def\Sesc{\Sigma_{\rm esc}}
\def\Srad{\Sigma_{\rm rad}}

\def\Dsolar{${\cal D}/\dsun$}
\def\Zsolar{$Z/\zsun$}
\def\DDsolar{\left( {{\cal D}\over \dsun} \right)}
\def\ZZsolar{\left( {Z \over \zsun} \right)}
\def\kms{{\rm km\,s}^{-1}\,}
\def\skms{$\sigma_{\rm kms}\,$}

\def\Scii{$\Sigma_{\rm [CII]}$}
\def\Sciimax{$\Sigma_{\rm [CII]}^{\rm max}$}
\def\CII{\hbox{[C~$\scriptstyle\rm II $]~}}
\def\CIII{\hbox{C~$\scriptstyle\rm III $]~}}
\def\OII{\hbox{[O~$\scriptstyle\rm II $]~}}
\def\OIII{\hbox{[O~$\scriptstyle\rm III $]~}}
\def\HH{\hbox{H$_2$}~} 
\def\HI{\hbox{H~$\scriptstyle\rm I\ $}} 
\def\HII{\hbox{H~$\scriptstyle\rm II\ $}} 
\def\CIion{\hbox{C~$\scriptstyle\rm I $~}}
\def\CIIion{\hbox{C~$\scriptstyle\rm II $~}}
\def\CIIIion{\hbox{C~$\scriptstyle\rm III $~}}
\def\CIVion{\hbox{C~$\scriptstyle\rm IV $~}}
\def\nhh{n_{\rm H2}}
\def\nhi{n_{\rm HI}}
\def\nhii{n_{\rm HII}}
\def\fhh{x_{\rm H2}}
\def\fhi{x_{\rm HI}}
\def\fhii{x_{\rm HII}}
\def\fd{f^*_{\rm diss}} 
\def\ks{\kappa_{\rm s}}

\def\cyan{\color{cyan}}
\definecolor{apcolor}{HTML}{b3003b}
\definecolor{afcolor}{HTML}{800080}
\definecolor{lvcolor}{HTML}{DF7401}
\definecolor{mdcolor}{HTML}{01abdf} 
\definecolor{cbcolor}{HTML}{ff0000}
\definecolor{sccolor}{HTML}{cc5500} 
\definecolor{sgcolor}{HTML}{00cc7a}